\documentclass[fleqn,10pt]{revtex4}
\usepackage{graphicx,amsmath,amssymb,mathrsfs}
\usepackage{epsfig}
\usepackage[colorlinks]{hyperref}

\begin{document}

\title{Generalized statistical mechanics of cosmic rays: Application to positron-electron spectral indices}

\author{G.\ Cigdem Yalcin$\,^1$ and Christian Beck$\,^{2*}$}

\affiliation{$\,^1$ Department of Physics, Istanbul University, 34134, Vezneciler,
Istanbul, Turkey \\
$\,^2$ School of Mathematical Sciences, Queen Mary University of London, Mile End Road, London E1 4NS, UK\\
$\,$ \\
$\color{blue} \,^*$ correspondence to c.beck@qmul.ac.uk}

\begin{abstract}
{\bf Cosmic ray energy spectra exhibit power law distributions over many orders of magnitude that are
very well described by the predictions of $q$-generalized statistical mechanics,
based on a $q$-generalized Hagedorn theory
for transverse momentum spectra and hard QCD scattering processes. QCD at largest center of mass energies predicts the entropic index to be $q=\frac{13}{11}$. Here we show that the escort duality of the nonextensive
thermodynamic formalism predicts an energy split of effective temperature given by
$\Delta kT =\pm \frac{1}{10} kT_H \approx \pm 18 $ MeV, where $T_H$ is the Hagedorn temperature.
We carefully analyse the measured data of the AMS-02 collaboration and provide evidence that the predicted temperature split is indeed observed,
leading to a different energy dependence of the $e^+$ and $e^-$ spectral indices.
We also observe a distinguished energy scale $E^* \approx 50$ GeV where the $e^+$ and $e^-$ spectral indices
differ the most.
  Linear combinations of the escort and non-escort $q$-generalized canonical distributions
   yield excellent agreement with the measured AMS-02 data in the entire energy range.}

\end{abstract}

\maketitle

\section{Introduction}

Statistical mechanics is a universal formalism based on the maximization of the Boltzmann-Gibbs-Shannon entropy
subject to suitable constraints. Despite its universal validity and success for short-range equilibrium systems, the applicability of the Boltzmann-Gibbs
formalism has severe restrictions: It is not valid for nonequilibrium systems, it is not valid for systems with
long-range interactions (such as gravity), and it is not valid for systems with a very small volume and fluctuating temperature (as probed
in scattering processes of cosmic ray particles at very high energies).
For these types of complex systems it is useful to generalize the formalism to a more general setting, based on the
maximization of more general entropy measures which contain the Shannon entropy as a special case.
Probably the most popular one of these generalizations is based on $q$-entropy (or Tsallis entropy), which leads
to power law distributions (the so-called $q$-exponentials),
but other generalized entropic approaches are possible as well \cite{tsallis1,tsallis-book,beck-review,turner1,turner2, turner3,q-duality,tirnakli}.
In high energy physics, a recent success of the $q$-generalized approach is that excellent fits
of measured transverse momentum spectra in high energy scattering experiments have been obtained \cite{wong},
 based on an extension of Hagedorn's theory \cite{hagedorn1,rafaelski,hage-temp}
to a $q$-generalized version and a generalized thermodynamic theory
  \cite{hagedorn2,hage-early,beck2000,new1,new2,new3,new4}. This includes recent experiments in the TeV region for $pp$ and $p\bar{p}$ collisions
  \cite{wong,cleymans,wilk,alice,exp1,exp2,exp3}, but there is also early work on
cosmic ray spectra \cite{tsallis-cosmic,beck-cosmic} and $e^+e^-$ annihilation \cite{bediaga,beck2000,beck2009}.


In this paper we systematically investigate the relevant degrees of freedom
of the $q$-generalized statistical mechanics formalism at highest center of mass energies and develop an effective theory of energy spectra of cosmic rays,
which are produced by scattering processes at extremely high energies (e.g. supernovae explosions). Some remarkable
predictions come out of the formalism in its full generality. First of all, the parameter $q$
is not an arbitrary fitting parameter but fixed at highest center of mass energies---
by comparing the predicted power law decay for transverse momentum
spectra with the hard QCD parton scattering amplitude \cite{wong}. In the parametrization we use in
the current paper this leads to $q=\frac{13}{11}=1.1818$. Next, the average temperature parameter for QCD scattering processes and subsequent hadronization is theoretically expected to
be the Hagedorn temperature, it cannot be varied either but is fixed by theory.
In fact, in this paper we will show that in full generality the $q$-generalized statistical mechanics
predicts a small temperature splitting close to the Hagedorn temperature,
by applying the escort and non-escort formalism of the theory \cite{BS,curado} (a manifestation of
the so-called $q$-duality of the generalized statistical mechanics \cite{turner1,turner2,turner3,q-duality,yalcin, tsallis2017}). Remarkably,
this temperature split prediction of the theory is confirmed by the recent AMS-02 measurements \cite{AMS1-1,AMS1-2,AMS1-3,AMS1-4, AMS2} of fluxes for cosmic ray electrons and positrons,
as we will show in this paper. The small temperature split leads to a different behavior of electron and positron
spectral indices as a function of energy, which so far was not understood, but is here identified as a feature
of the generalized statistical mechanics formalism.

{
\section{Background and previous work}

 Let us motivate the use of generalized statistical mechanics as an effective theory for cosmic rays in a bit more detail.
Generally, it is
well known that a large variety of different physical processes contribute to
the observed spectrum of cosmic rays at the Earth. First of all, there is the primary statistics generated by the
primary production and acceleration process of cosmic rays in  sources such as supernovae, active galactic nuclei, heavy stars with stellar winds, or even the big bang itself.
Not much is known about the primary production process, but of course QCD scattering processes at
extremely high energies will play an important role.
The primary spectrum is then modified by subsequent propagation through the interstellar medium, { where
magnetic turbulence, ionization, Coulomb interactions, Bremsstrahlung, inverse Compton and synchrotron
processes play an important role,
as well as fragmentation in the interstellar medium}. There is not just one process
 but a whole spectrum of many different complex processes involved,
 which suggests the use of generalized statistical mechanics as an effective
 theory. Finally, when entering the heliosphere
 the cosmic ray flux is further modulated by solar activity \cite{solar1,solar2}.

 QCD scattering and selfsimilar hadronic fragmentation processes have been previously successfully described by $q \not= 1$ theories \cite{wong,new1,new2,new3,new4},
 and applied to data taken at LHC \cite{exp1,exp2,exp3}.
    It is precisely these types of scattering processes that we postulate as main ingredients to imprint
   on the momentum spectrum of primary cosmic rays, an initial spectrum set by QCD that in its main features is basically conserved, while other important processes then only yield small
   perturbations of the initial spectrum set by QCD, at least in a statistical sense.
Still subtle differences between electron and positron statistics
are possible, as well as deviations from simple QCD behavior, which we will discuss in
more detail in the following sections.


Among phenomenological approaches to understand high energy cascading scattering processes,
power laws associated with Tsallis statistics are by now widely used
by many groups \cite{wong,cleymans,wilk,new1,new2,new3,new4}. In fact they yield
surprisingly good fits of a variety of data sets for different systems. This approach uses the
assumption that the highly excited and fragmented states formed in high energy
collisions follow Tsallis statistics instead of Boltzmann statistics. Initially only regarded
as a mathematical playground for more general versions of statistical mechanics, based on
the maximization of more general
entropy measures \cite{tsallis1,tsallis-book,beck-review,q-duality}, the approach has more recently led to more sophisticated
theories which do produce excellent agreement with
experimental data, much more beyond the original fitting approach. Recent models are now going much more in-depth on which specific type of formalism is appropriate, and relate the entropic index to QCD scattering processes in the perturbative regime \cite{wong,new3}.
But also the nonperturbative regime is accessible, where the generalized statistical mechanics formalism arises
out of the self-similarity of the fragmentation process and can be related to the fractal-like
structure of hadrons within the MIT bag model for hadron structure \cite{new1,new2}. Moreover,
Tsallis statistics is generally well-known to be highly relevant for velocity distributions in astrophysical plasmas
(here these distributions come under the name {\em Kappa-distributions}, see, e.g.,
   \cite{livadiotis} for a recent review). Importantly, an entropic index $q \not= 1$  can also arise from nonequilibrium dynamics, in particular from suitable spatio-temporal fluctuations of an
intensive parameter such as the local (inverse) temperature (the superstatistics approach \cite{beck-cohen}, as
 developed by Beck and Cohen (see note in the acknowledgements)).
 In this paper, for the very first time, we apply these techniques borrowed from generalized statistical mechanics to analyse
  the AMS-02 data sets.}

\section{Results}

\subsection*{$q$-dualities and theoretical prediction of a temperature split}

{While there are by now many different versions of the nonextensive formalism in high energy physics,
each describing different aspects of different scattering systems, for cosmic rays consisting of electrons
 and positrons we start from one of
the simplest versions:}
As in early work on cosmic ray data \cite{tsallis-cosmic, beck-cosmic} we start from $q$-generalized canonical distributions of the form
\begin{equation}
p(E)\sim \frac{E^2}{(1+(q-1)\beta_0 E)^\frac{1}{q-1}}.
\label{1}
\end{equation}
Here $q$ is the entropic index and $\beta_0$ is an inverse temperature parameter. $E$ is the energy of the particle.
For $q \rightarrow 1$ the ordinary Maxwell Boltzmann distribution

\begin{equation}
p(E)\sim {E^2}e^{-\beta_0 E}
\label{xxx}
\end{equation}
 is recovered. The above distributions are $q$-generalized canonical distributions in the nonextensive formalism,
$E^2$ is a phase space factor. They maximize Tsallis entropy $S_q=\frac{1}{q-1} \sum_i (1-p_i^q)$ subject to suitable constraints \cite{tsallis1, tsallis-book, beck-review, q-duality, curado}.
They generate power law distributions for large energies $E$ and are thus very well suited to fit power law distributions observed
in cosmic ray physics. Defining the spectral index $\gamma$ by $\gamma := d \log p(E)/ d \log E$ we obtain the result
that a single distribution of the form eq.~(\ref{1}) generates the spectral index
\begin{equation}
\gamma (E) =2-\frac{\beta_0 E}{1+(q-1)\beta_0 E} \to 2 - \frac{1}{q-1} \;\;\;\;\;\;\;\;\; (\beta_0 E \to \infty)
\end{equation}

A physical motivation for the occurrence of the above asymptotic power-law distributions
   can be given by the superstatistics approach, which quite generally describes
   the effect of big temperature fluctuations
 in nonequilibrium situations \cite{beck-cohen}. We can write eq.~(\ref{1})
 in the equivalent form
\begin{equation}
 \int_0^\infty d \beta f(\beta) E^2 e^{-\beta E}
=\frac{E^2}{(1+(q-1)\beta_0 E)^{1/(q-1)}}
\end{equation}
where $f(\beta)$
is a $\chi^2$ distribution with $N=2/(q-1)$ degrees of
freedom and $\beta_0= \int_0^\infty d\beta \, \beta f(\beta)=\langle \beta \rangle$ is the average of
a fluctuating random variable $\beta$ that is distributed with $f(\beta)$. One can easily check that $q=\langle \beta^2 \rangle/\beta_0^2$,
so $q-1$ is a measure of the width of the inverse temperature fluctuations
\cite{wilk2000, beck2001}. The physical interpretation is that power law Boltzmann
factors $(1+(q-1)\beta_0 E)^{-1/(q-1)}$ arise from ordinary Boltzmann factors $e^{-\beta E}$ in nonequilibrium
situations where there is a distribution $f(\beta)$ of inverse temperatures $\beta$, after integrating over
all possible $\beta$ weighted with $f(\beta)$. The relevance of
      temperature fluctuations in cosmic ray physics
      has been previously emphasized in \cite{beck-cosmic,beck2009}.

Our main goal in the following is to derive physically plausible values for $q$ and $\beta_0$ for
cosmic ray energy spectra, in particular for $e^+$ and $e^-$ cosmic rays,
and to proceed to linear combinations of distributions of type (\ref{1}).
To this end we assume that the underlying primary production process of the cosmic rays is QCD parton scattering at largest possible energies. Essentially protons (which make up the dominant component of cosmic rays) collide at extremely high energies
and produce other particles in the process. In these high energy scattering processes and subsequent
hadronisation cascades a large number of baryons and anti-baryons
as well as mesons are produced. These hadrons ultimately decay to stable particles, including electrons and positrons, which
continue their way as cosmic rays, but imprinted on
their statistical transverse momentum distribution is the original high-energy scattering and
hadronisation process that took place at
the Hagedorn temperature.

Wong et al. \cite{wong} have derived that for hard parton  QCD scattering processes the asymptotic
power law dependence of the cross section imposed by the leading QCD scattering amplitude is
$E \frac{d^3 \sigma}{d p^3} \sim E^{-\frac{9}{2}}$ (see their eqs. (45) and (46) in \cite{wong}) which in our parametrization using the phase space factor $E^2$ implies
$
p(E)\sim{E^{-\frac{7}{2}}}.
$
Thus QCD at largest energies fixes in our formula (1) the parameter $q$ to be
\begin{equation}
2-\frac{1}{q-1}=-\frac{7}{2}\Leftrightarrow q=\frac{13}{11}=1.1818
\label{4}
\end{equation}

 For QCD scattering processes, the average temperature $1/\beta_0$ should be essentially given by the Hagedorn temperature. To get the precise relation, we recall that
in the nonextensive formalism there is a second important canonical distribution, the so-called escort distribution \cite{BS, curado}.
This is given by
\begin{equation}
\hat{p}(E)\sim E^2\frac{1}{(1+(\hat{q}-1)\hat{\beta_0}E)^\frac{\hat{q}}{\hat{q}-1}}
\label{5} \end{equation}
where $\hat{q}$  and $\hat{\beta_0}$ are the entropic index and inverse temperature parameter for the escort distribution. Basically, the escort distribution is obtained by raising all given microstate probabilities $p_i$ to the power $q$
and renormalizing this distribution (see \cite{BS, curado} for
details).
In this way, for any $q\not= 1$ two degrees of freedoms arise out of the generalized statistical mechanics treatment:
Escort distributions and non-escort distributions.
Both formalisms can be mapped onto each other, this is called the {\em escort duality}. The escort duality
can also be understood in terms of superstatistics,
a related concept was called type-A and type-B superstatistics in \cite{beck-cohen}.
There are actually two $q$-dualities in
the nonextensive formalism, corresponding to the replacements $q \to 2-q$ and $q\to 1/q$.
These are sometimes called {\em additive} and {\em multiplicative} duality \cite{q-duality,tsallis2017}, and they can be combined to
give the {\em escort} duality $q \to 2-\frac{1}{q}$.

In our case, QCD fixes the power law index
\begin{equation}
n:=\frac{1}{q-1}=\frac{\hat{q}}{\hat{q}-1}
\label{6}
\end{equation}
to $n=\frac{11}{2}$ at largest energies. From this we derive
a relation between $q$ and $\hat{q}$:
\begin{equation}
q=2-\frac{1}{\hat{q}}  \Leftrightarrow  \hat{q}=\frac{1}{2-q}
\label{7}
\end{equation}
which is just the escort duality. The distributions
$p(E)$ given in eq.~(\ref{1}) and $\hat{p}(E)$ given in eq.~(\ref{5}) describe the same physics and should be the same, no matter whether we mathematically use the escort or non-escort
formalism. This yields a relation between the inverse temperature parameters $\beta_0$ and $\hat{\beta_0}$:
\begin{equation}
(q-1)\beta_0=(\hat{q}-1)\hat{\beta_0},
\label{8}
\end{equation}
which can be solved to give
$
\hat{\beta_0}=\frac{q-1}{\hat{q}-1}\beta_0 =(2-q)\beta_0
$.
In terms of average temperature $kT=\beta_0^{-1}$ the fixed index $n$ implies the existence
of two different temperature parameters:
\begin{equation}
\hat{T}=\frac{1}{2-q}T
\label{10} \end{equation}

If we define the Hagedorn temperature $T_H$ as the average of these two degrees of freedom,

\begin{equation}
T_H=\frac{1}{2}(\hat{T}+T)
\label{15} \end{equation}
we obtain
$
T_H=\frac{1}{2}(\frac{1}{2-q}+1)T
=\frac{1}{2}  \frac{3-q}{2-q}T
$.
It then follows
$
T=2  \frac{2-q}{3-q}T_H,
$
and
$
\hat{T}= \frac{2}{3-q}T_H
$
so that $\Delta T :=\hat{T}-T_H= \frac{q-1}{3-q}T_H$.
In particular, for the QCD hard scattering processes with $n=\frac{11}{2}$ we have
$q=\frac{13}{11}=1.1818$, $\hat{q}=\frac{11}{9}=1.2222$ and
the predicted split in temperature is given by
\begin{equation}
T=\frac{9}{10}T_H=T_H-\frac{1}{10}T_H,
\label{21} \end{equation}

\begin{equation}
\hat{T}=\frac{11}{10}T_H=T_H+\frac{1}{10}T_H.
\label{22} \end{equation}
This means that two different effective  temperatures degrees of freedom come out of the formalism,
and this is an experimentally testable prediction that one can check on the measured cosmic ray data.
There are slight uncertainties in the knowledge of the precise value of the Hagedorn temperature in the literature, for
our fits in the following we
use the value $kT_H=$ 180 MeV as in \cite{hage-temp}, getting $kT$=162 MeV and $k\hat{T}$=198 MeV.

{The main idea of our derivation of the above temperature split is that in the $q$-generalized formalism the physical meaning is
not directly connected with the entropic index $q$ but with the power law exponent $n$,
since the latter one is directly measurable. However, it does not matter if one writes $n$ in either the form
$n=1/(q-1)$ or $n=q/(q-1)$. But if one does both then the temperature must be different in each
of these cases,
as shown above. This temperature splitting occurs only for $q\not= 1$, whereas ordinary statistical mechanics with $q=1$
has a temperature fixed point}.

\subsection*{Comparison with AMS-02 measurements}

Let us now compare our theoretical predictions with the measurements of the AMS-02 collaboration \cite{AMS1-1,AMS1-2,AMS1-3,AMS1-4,AMS2}. Fig.~1 shows that formula (1) with the predicted $q$-value $q=\frac{13}{11}$ and the two predicted temperatures $T$=$T_H$$\pm\frac{1}{10}T_H$ yields an excellent fit of the AMS data up to energies of about 50 GeV. Remarkably, electrons are described by $\hat{T}$=$\frac{11}{10}T_H$ and positrons by $T$=$\frac{9}{10}T_H$, thus giving physical meaning to the
two degrees of freedom that come out of the
generalized statistical mechanics theory. The fits
are very sensitive to the $q$-values used, up to 3-4 digits of precision in $q$ can be distinguished,
confirming the QCD value $q=1.1818$ derived from theory, as well as the predicted temperature split.

{The error bars of the data that are plotted in Fig.~1 (as listed by AMS-02) are small, when these error
             bars are plotted in our logarithmic plot,
             they are of the same order of magnitude as the size of the symbols, and have therefore been
              omitted in our Fig.~1 for reasons of better visibility.}
Note that in Fig.~1 all parameters used are predicted, and not fitted. Excellent agreement is obtained
up to energies of about 50 GeV. The only fitting parameters
used in Fig.~1 is the absolute strength of the flux (corresponding to the vertical position of the curves in the logarithmic plots):
The electron flux amplitude $A_-=770$ is about 7 times stronger than the positron amplitude $A_+=115$.

{Our $q$-generalized statistical mechanics has no tool to predict the amplitude parameters $A_\pm$ from first principles, it can only give predictions on the shape of the spectrum, given some fixed amplitude parameter for a given particle species.
The conventional wisdom to explain why $A_- >> A_+$ is that electrons are mainly produced in the
primary acceleration process, whereas positrons are believed to be mainly secondary particles arising from collisions
with the interstellar plasma.
An alternative approach, more in line
with what we see from the actual fits of the $e^+$ and $e^-$ spectra, would be that both positrons and electrons are
initially produced in the same primary QCD cascading process with similar amplitude (and with the same $q$-value
and only a slightly different Hagedorn temperature), but that then
an energy-independent absorption process
sets in where anti-particles (positrons) have a higher probability of being absorbed than particles (electrons), leading finally to the observed ratio $A_+/A_- \approx 1/7$.
One might speculate that this absorption process has
to do
with the (unknown) process that leads to CP symmetry breaking in the universe and the dominance of matter
over anti-matter abundance. As said, a full understanding of the amplitude ratio $A_+/A_-$ is out of reach of our simple
statistical mechanics model, but has been the subject of other papers.}

{Another important feature to discuss, important for the low-energy end of the spectrum, are solar modulation effects.
                  At low energies (typically $E \leq 1$ GeV) the flux of $e^+$ and $e^-$ varies in time due to the changing pattern of solar activity,
              thus making the measured flux of electron and positron cosmic rays time dependent.
              Solar modulations occur because of interactions of cosmic rays entering the solar heliosphere;
              they are due to charge sign dependent propagation in the solar magnetic field. For a review, see
     \cite{solar1}. It is clear that our prediction, based on a simple statistical mechanics model which
     does not know about this effect, {cannot give quantitative predictions related to
     the solar modulation effects. What is measured in the experimental data is the time-integrated effect, but this effect
     is not zero but always depletes the spectrum at very low energies. What our analysis and the good agreement with the data seems to indicate is that the averaged effect of solar modulation
     is negligible at energies down to around 1 GeV, at least in
     the logarithmic scale
     that is being used in Fig.~1.}

                          Let us use previous work to estimate the order of magnitude of the expected solar periodic modulation. Fig.~20 of \cite{solar2} (based on PAMELA data) indicates that at an energy of 0.7 GeV, the lowest energy considered in our Fig.~1, the half-yearly measured flow changes by a factor of up to 1.5,
due to the solar modulation effect. Although the effect is highly significant for a detailed
understanding of the time dependence of the low-energy spectra, a factor of 1.5 is hardly visible
in the logarithmic plot of our Fig.~1, where it just leads to a shift of the data that has the same order of magnitude as the symbol size, given the logarithmic scale of the figure. That said, an interesting future project would be a precision
comparison of the time-dependent fitting parameters of the generalized statistical mechanics approach
with time-dependent data as presented e.g.\ in \cite{solar2}.
}

\begin{figure}[h]

\vspace{2cm}
\includegraphics[width=9cm]{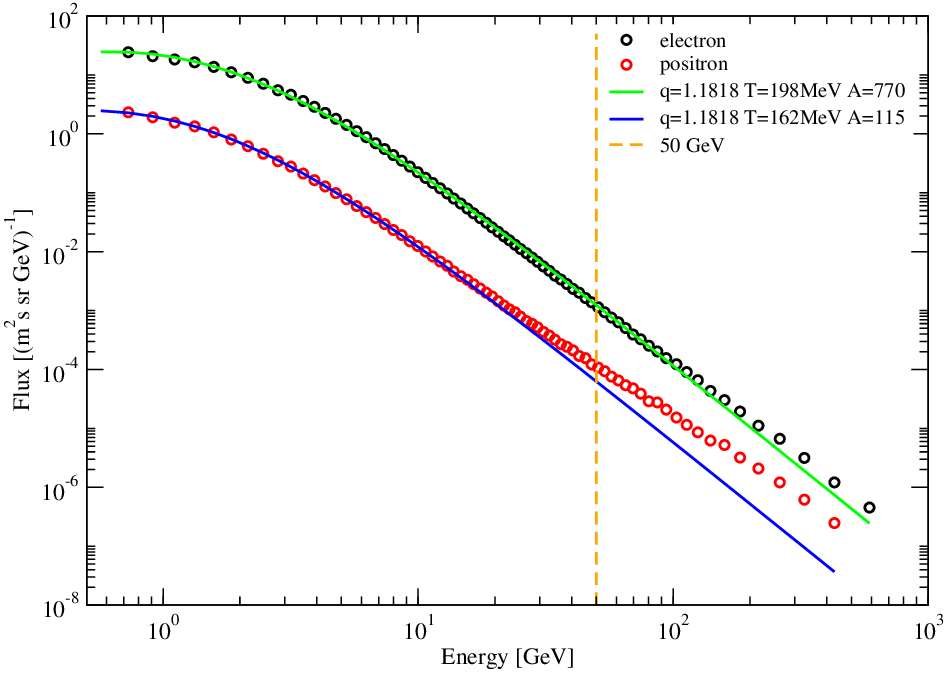}
\caption {Flux $\Phi (E)$ of $e^+$ and $e^-$ primary cosmic ray particles of energy $E$ as measured by AMS-02 \cite{AMS1-1,AMS1-2,AMS1-3,AMS1-4,AMS2} and theoretical prediction of the $q$-generalized Hagedorn theory (solid lines).}

$\;$

\end{figure}

Let us now proceed to higher energies.
In Fig.~2 it is shown that at an energy of the order 50 GeV our formula
based on a single $q$-exponential as given in eq.~(\ref{1}) starts to deviate significantly from the measured flux data. {Positrons start to deviate earlier than electrons. We will later give a suitable
definition of a joint transition point by analysing the ratio of local positron and electron
spectral indices, which is an amplitude-independent quantity.}

\begin{figure}[h]
\vspace{2cm}
\includegraphics[width=9cm]{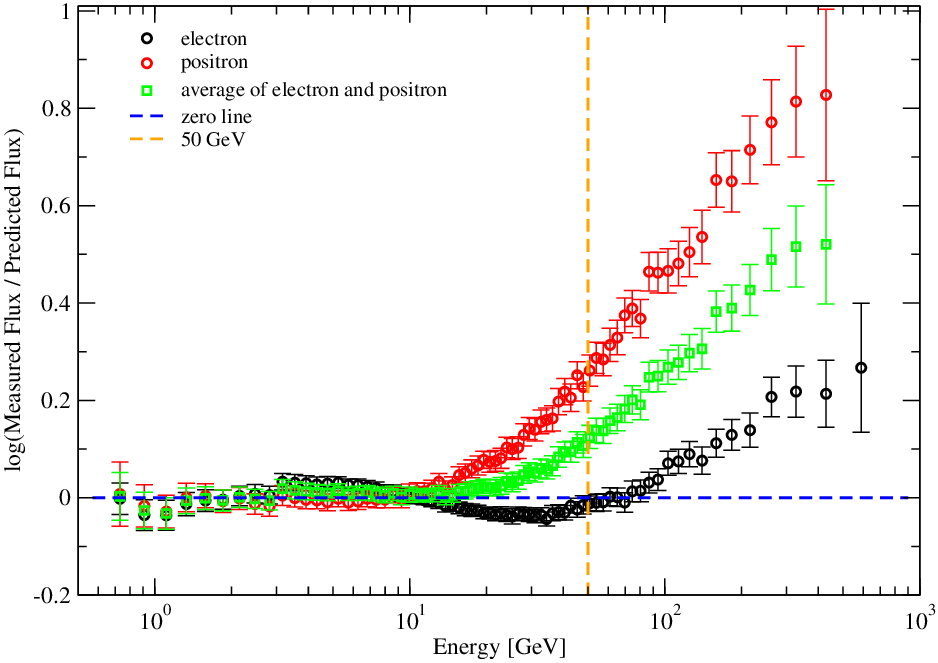}
\caption {Ratio of measured flux to predicted flux as given by eq.(1) as a function of energy $E$ of the cosmic ray particles. The data correspond to AMS-02 measurements
of electrons, positrons, {and both species together}. The vertical line indicates the energy $E$= 50 GeV. }
\end{figure}

Apparently, an additional process starts to contribute to the cosmic ray flux for energies $E$
larger than about 50 GeV. We found that this crossover is very well described by a linear combination of
 generalized canonical distributions where the entropic index takes on two values, namely the
 QCD value $13/11=1.1818$ and the escort value $11/9=1.2222$,
 evaluated at temperature $T$ for positrons and $\hat{T}$ for electrons.
            Fig.~3 shows that in the entire energy range the measured cosmic ray flux is very well fitted by the linear combination

\begin{equation}
\! \! \! \! P_\pm (E)=A_\pm (\frac{E^2}{(1+(q-1)\beta_0 E)^\frac{1}{q-1}}+C_\pm \frac{E^2}{(1+(\hat{q}-1)\beta_0 E)^\frac{1}{\hat{q}-1}})
\label{24} \end{equation}
where $C_+\approx$ 0.04 for positrons,  $C_-\approx$ 0.0053 for electrons, and  $\beta_0=T^{-1}$, respectively $ \hat{T}^{-1}$.

\begin{figure}[h]
\vspace{2cm}
\includegraphics[width=9cm]{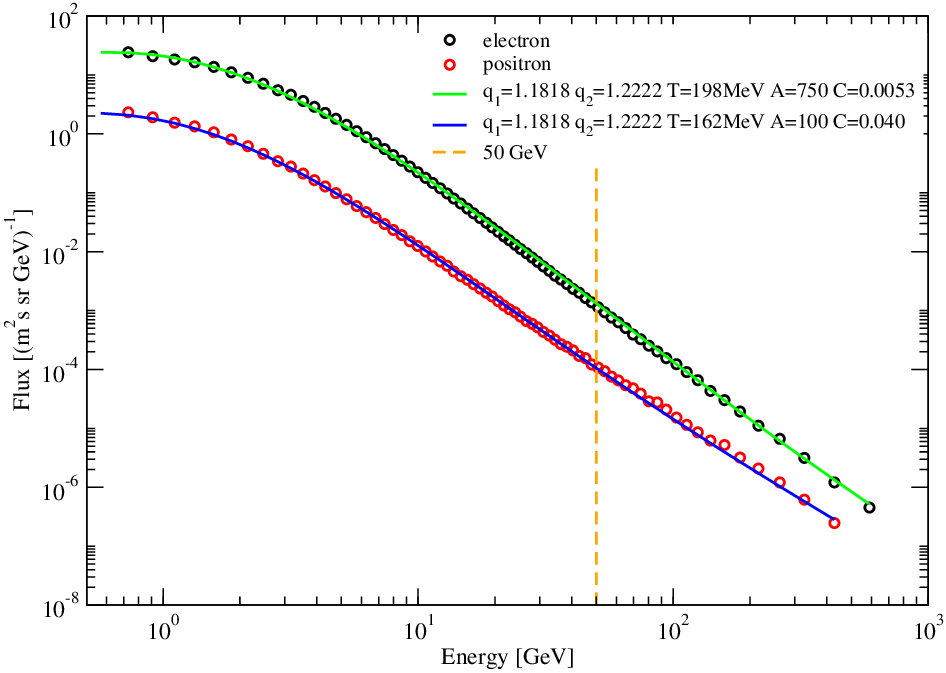}
\caption{The measured AMS-02 data are very well fitted by linear combination of escort and non-escort distributions (solid lines).}
\end{figure}

Apparently, the power law exponent $-\frac{1}{\hat{q}-1}=-4.5$ is larger than the power law exponent
$-\frac{1}{q-1}=-5.5$, and hence the former term dominates the large energy behavior if $E\rightarrow\infty$. The crossover scale observed in the AMS data is roughly 50 GeV.
We may physically interpret this crossover again as an effect of the two degrees of freedom (escort and non-escort) that are predicted by the $q$-generalized statistical mechanics formalism: The physically relevant generalized canonical distribution is a mixed state of both, and hence at smaller energies the behavior is dominated by the power law exponent $\frac{1}{q-1}$, whereas at higher energies it is dominated by $\frac{1}{\hat{q}-1}$.

{Let us summarize the main assumptions underlying our approach. The generalized statistical mechanics presented in this paper
               gives concrete predictions on the shape of the flux spectrum as a function of observed energy of the particles, but it relies on the assumption that mainly QCD scattering and fragmentation process determine the shape of the spectrum, whereas other processes are either neglected or absorbed in terms of an effective description.
               While we do predict general spectral shapes based on
generalized statistical mechanics methods, we certainly cannot give predictions on the size of the absolute flow of particles as modified by the local astrophysical environment. In particular, very low energy
spectra of electrons in the MeV region (such as those discussed in \cite{add1}) are certainly out of
reach of the current formalism.

The generalized statistical mechanics model applies to both primary and secondary production processes of
electrons and positrons, just the relevant parameter $q$
 is different, depending on the center of mass energy $E_{CMS}$ considered.
 In particle physics scattering processes at typical LHC center of mass energies $E_{CMS}$
in the TeV region, $q$ is
observed to be typically around $q \approx 1.10$ \cite{new3,wong,add3}, and one has theoretical predictions that for
$E_{CMS} \to \infty$, $q$ should approach 1.22 \cite{beck2000}. The latter value fits much
 better the observed cosmic ray spectra, hence
our conjecture is that both electron and positron spectra get their $q \approx 1.2$ imprinted already
in the primary production process, at significantly higher energies than the
TeV region. The secondary scattering processes then have a $q$ more in the region
like the LHC data, $q\approx 1.1$, so this secondary power law decays stronger and can thus be neglected
as compared to the primary power spectrum which decays slower since $q$ is bigger. Generally,
the power laws decay as $E^{2-\frac{1}{q-1}}$, so the largest $q$ dominates the relevant spectrum.
In this context it is interesting to note that at very large center of mass
energies the existence of both a limiting $q$
and a limiting $\beta$ has been predicted on the basis of thermodynamic consistency relations\cite{add2}.
A systematic comparison
   with LHC scattering data for various particle species was presented in \cite{add3}.}

\subsection*{WIMP annihilation and other new physics}

An observed crossover feature in the cosmic ray flux pattern
of particles and anti-particles such as the one in Fig.~3 can have many origins, ranging from conventional astrophysical explanations to
more speculative explanations such as dark matter  annihilations and/or decay.
Let us here concentrate on a possible interpretation
in terms of WIMP physics, following the ideas presented in \cite{hooper1,hooper2,hooper3,aachen,conrad,prl1,prl2}.
    If there is a weakly interacting massive particle (WIMP) underlying dark matter, then WIMP annihilation
    and/or  WIMP decay can
      modify the relative abundance of $e^+$ and $e^-$ flux. In particular, a distinct change of behavior of the flux is expected at a
       threshold energy given by the WIMP mass.
                         Presently WIMP masses of $\sim 50$ GeV near the $Z^{\circ}$ pole are still consistent with particle collider experiments
in a variety of models and have been proposed as an explanation for the observed
$\gamma$-ray excess from the center of the galaxy \cite{hooper1,aachen,hooper2,hooper3, conrad}.
Also, a recent analysis of the AMS-02 antiproton flux data appears to be consistent with a WIMP
mass of about 50 GeV \cite{prl1,prl2}.
Interestingly, our fit of the electron positron data seems to indicate $A_-C_-=A_+C_+$, meaning the excess flux contribution starting to dominate in eq.~(\ref{24}) from
50 GeV onwards has the same amplitude for electrons and positrons.

To better characterize the crossover point,
we define it as the point where the spectral indices $\gamma_+$ and $\gamma_-$ of positrons and electrons differ the most: $|\gamma_+-\gamma_-|=max$ implies that $\gamma_+/\gamma_-$ has a minimum.
 Fig.~4 shows the ratio $\gamma_{+}(E)/\gamma_{-}(E)$ as a function of energy $E$ as measured by AMS-02 \cite{AMS2} ({the error bars of the ratio were estimated by standard methods}).
We observe that there is a local minimum at $E=E^*=(50 \pm 10)$ GeV,
at this special point we have  $\gamma_+-\gamma_-=\frac{1}{2}$ and
$\gamma_+/\gamma_-=0.8462=11/13=q^{-1}$ (Fig.~3 in \cite{AMS2} indicates $\gamma_+=-2.75$ and $\gamma_-=-3.25$).
It is remarkable that the unknown crossover process taking place at $E^*$
satisfies
$
\gamma_+(E^*)=\hat{\gamma} -\frac{1}{4}$ and $ \gamma_-(E^*)=\gamma +\frac{1}{4}
$,
so there is an antisymmetric correction $\pm \frac{1}{4}$ to
 the scaling exponents $\gamma=2-\frac{1}{q-1}$ and $ \hat{\gamma}=2-\frac{1}{\hat{q}-1}$ associated with $q=13/11$ and $\hat{q}=11/9$. Fig.~4 indicates that over a wide range of energies centered around
  $E^*=50$ GeV the ratio $\gamma_+/\gamma_-$ exhibits a quadratic logarithmic energy dependence,
 \begin{equation}
 \frac{\gamma_+}{\gamma_-}(E)=q^{-1}+C \left( \ln \frac{E}{E^*} \right)^2 \label{para}
 \end{equation}
 where $q=13/11$, $C=0.037$. The minimum at $E^*$ corresponds to a distinguished superstatistical state
 satisfying $N_-=N_++1$, i.e. the $\chi^2$-distributions of inverse
 temperatures that are relevant for electrons and positrons differ by precisely 1 degree of freedom
 at this point. Physically this could be interpreted in terms of an additional particle degree of freedom
 that does influence the temperature fluctuations and is seen by electrons, but not by positrons.


$\;$

$\;$

\begin{figure}[h]

\includegraphics[width=9cm]{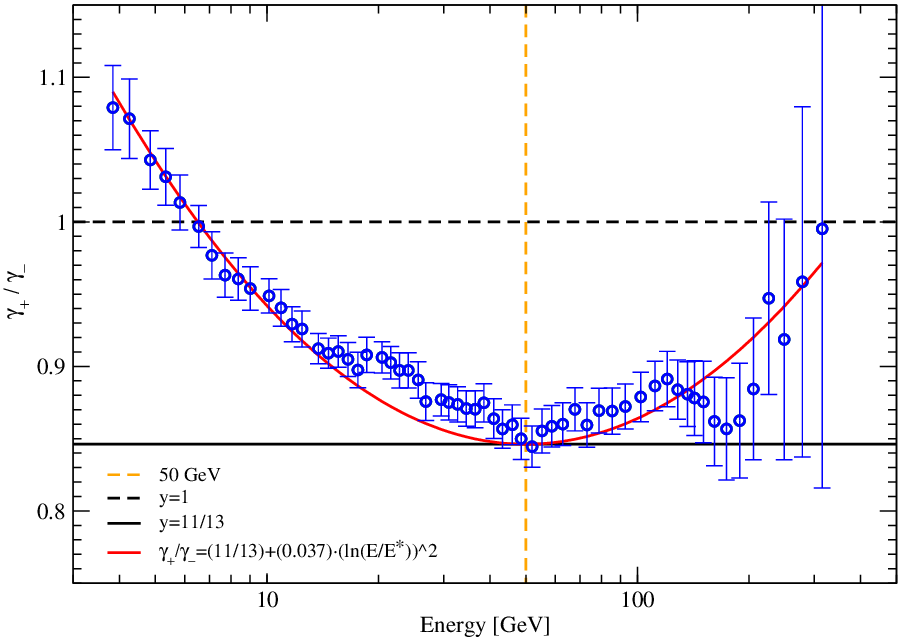}
\caption{Ratio $\frac{\gamma_{+}}{\gamma_{-}}$ of positron and electron spectral indices as a function of energy $E$ (data from
the AMS-02 collaboration \cite{AMS2}).
The curve $\frac{\gamma_{+}}{\gamma_{-}} $ has a minimum at $E=E^*= (50 \pm 10)$ GeV and is well fitted by
eq.~(\ref{para}).}
\end{figure}

Note that QCD on its own (without any $q$-generalized statistical mechanics) predicts the same spectral indices $\gamma_+= \gamma_-$ for positrons and
electrons. But the AMS-02 data (and previous data from PAMELA \cite{pamela}) have shown that there is a significant
difference between spectral indices of {electrons and positrons}, an asymmetry that is not fully
understood so far. Whatever the theoretical explanation,
the measured difference $|\gamma_+- \gamma_-|(E)$ as a function of energy can be regarded as a measure of the
significance of the new physics underlying this difference.
We see from the data analysed in Fig.~4 that the deviation is strongest at $E=E^* \approx 50$ GeV.
This observation is based on the measured AMS-02 data and is independent of any model assumption.
It is intriguing to notice that the deviation is strongest at a mass scale that is the same as the extracted WIMP
mass for other data sets such as the $\gamma$ ray flux from the center of the galaxy and
the antiproton flux data \cite{prl1,prl2,hooper1,hooper2,hooper3}.

{Summarizing, our analysis
of the very precise AMS-02 data has shown that for cosmic rays with an energy below the order of magnitude 50 GeV the observed statistics of both
electrons and positrons is well described by what one would expect from QCD at largest possible center of mass energies
$E_{CMS} \to \infty$,
a $q$-statistics with $q=13/11$ (equivalent to a power law decay with exponent -5.5), and a temperature given by the Hagedorn temperature but with
a small split in temperature, as expected from the $q$-generalized formalism in its full generality. While
there are certainly important further processes for electrons and
positrons on their way to Earth, such as e.g. energy loss by inverse Compton and synchrotron processes,
our data analysis suggests that these additional processes do not significantly
modify the primary entropic index $q$ set by QCD
at largest center of mass scattering energies.
Indeed generalized statistical mechanics with $q \not= 1$
has been shown in many previous papers to be relevant for QCD and hadronization cascades, but less so
for electromagnetic interaction processes which are better embedded into ordinary $q=1$
statistical mechanics. So our main hypothesis, supported by the data, is that the primary
entropic index $q\not=1 $ set by QCD is essentially conserved, at least in a statistical sense,
even if many further electrodynamical processes accompany the individual
electrons and positrons on their way to Earth.
But QCD alone cannot explain the spectra measured by AMS-02 in the high-energy region: What we actually observe is a transition point around $E^* \approx 50$ GeV where
the observed power law decay exponent switches from $-5.5$ to $-4.5$, and where one needs an additional mechanism to explain the excess
of electrons and positrons as compared to the original QCD prediction.
As WIMPs can decay into $e^+e^-$ it is
natural to associate this observed increased flux to some new physics associated with WIMP decay.
However, very recent experiments seem to provide growing evidence that WIMPs might not exist at all
\cite{wimp1,wimp2}.
In this case a more conservative interpretation of the transition would be that both
escort and non-escort distributions are realized in QCD scattering processes,
with a crossover scale $E^*$ where the escort index $-4.5$ starts to dominate the behavior.}

\section{Discussion}

In this paper we have applied $q$-generalized statistical mechanics methods
to high energy scattering processes, which lie at the root of the production process of cosmic rays.
These QCD scattering processes are thermodynamically described by the Hagedorn temperature, they are very
similar in their momentum characteristics to the collision processes of TeV protons that are
performed in experiments on the Earth \cite{wong,alice},
except that they can take place at much higher center of mass energy.
Stable particles
such as electrons and positrons finally arise out of the hadronisation cascade by decays of
pions, neutrons and other hadrons, and they
memorize the momentum statistics of the Hagedorn fire ball.
In this paper we carefully investigated the discrete degrees
of freedom that are contained in the $q$-generalized statistical mechanics
 describing this. We showed that there are basically two degrees of freedom
 in the formalism that correspond to escort and non-escort distributions, and which allowed us
 to identify different statistical behavior of particle and anti-particle degrees of freedom,
 as observed in the measured cosmic ray spectra.

 When comparing with the experimental data, the result of our analysis were two distinct energy scales:
A large energy scale of about $(50 \pm 10)$ GeV, where the spectral indices of $e^+$ and $e^-$ differ the most,
and a small energy scale of about $(18\pm 1)$ MeV, which corresponds to a splitting of the effective Hagedorn temperature.
The former energy scale sets the scale where the escort distribution power law starts to dominate the non-escort
power law. The latter energy leads to a slightly different temperature statistics for electrons and positrons.

 An interesting observation is that the theoretically predicted and experimentally observed split in Hagedorn temperature of
 $\frac{q-1}{3-q}kT_H \approx (18\pm 1)$ MeV  coincides with the mass scale of
  the recently postulated protophobic gauge boson
    \cite{feng1,feng2,experimental,kitahara,rose}. A physical interpretation could be that some of the kinetic energy represented  by the Hagedorn temperature is either absorbed or enhanced by this exotic postulated protophobic gauge boson should it exist. This would potentially be a quantum manifestation of the observed temperature split.
    By definition a protophobic gauge boson as introduced in \cite{feng1} couples to neutrons but not to protons. Assuming that electrons in
    cosmic rays arise to a significant part from $\beta$-decays of neutrons, then this could influence
    the observed momentum statistics and effectively lead to a slight asymmetry in the temperature statistics of electrons
    (which arise from neutron decays) as compared to that of positrons (which don't).
    While the existence of the protophobic
 gauge boson still requires independent experimental verification,
 it is interesting that the $q$-generalized Hagedorn theory allows for a possible embedding of these types
 of energy scales close to the Hagedorn temperature.

 To conclude,
 we have shown that the different energy dependence of the spectral indices
  of positron and electron cosmic rays is well explained by a $q$-generalized Hagedorn theory. The value of the parameter $q=13/11$ and the relevant temperature
   parameters were derived, they are not fitting parameters but theoretically determined and appear to be in excellent agreement with
 the measurements of the AMS-02 collaboration for $E< E^*$. The generalized statistical mechanics formalism together with the AMS-02 data analysed indicates the
 existence of two energy scales that could be potentially associated with new physics, a low-energy temperature split in Hagedorn temperature
 given by $\frac{1}{10} kT_H\simeq 18$ MeV and a crossover scale $E^*\simeq 50$ GeV where a process beyond QCD sets in, of
 unknown nature. These two energy scales
 appear to coincide with the mass scale of the recently postulated protophobic gauge boson \cite{feng1,feng2,experimental, kitahara,rose} and the mass scale of a WIMP that could explain the excess of $\gamma$ rays from the galactic center
 \cite{hooper1,aachen,hooper2,hooper3,conrad}.

\section*{Acknowledgements}
{We dedicate this
 paper to the memory of Prof. E.G.D. Cohen, Boltzmann medalist of the year 2004, who made many
 outstanding research contributions in nonequilibrium statistical mechanics and superstatistical approaches, and who passed away on
 24 September 2017.}

 G.C.Y. gratefully acknowledges the hospitality of Queen Mary University of London, School of Mathematical Sciences,
during a research visit in 2017. C.B.'s research is supported by EPSRC under grant number EP/N013492/1.



\end{document}